\documentclass[a4paper,10pt,twoside]{cpc-hepnp}

\usepackage{multicol}
\usepackage{graphicx}
\usepackage{booktabs}
\usepackage{amssymb,bm,mathrsfs,bbm,amscd}
\usepackage[tbtags]{amsmath}
\usepackage{lastpage}
\usepackage{subfigure}

\begin{document}

\fancyhead[c]{Submitted to 'Chinese Physics C'}


\title{Producing Terahertz Conherent Synchrotron Radiation Based On Hefei Light Source\thanks{Supported by National Natural Science
Foundation of China (11375176) }}

\author{%
      Xu De-Rong$^{1)}$\email{xuderong@mail.ustc.edu.cn}%
\quad Xu Hong-Liang$^{2)}$\email{hlxu@ustc.edu.cn}%
\quad Shao Yan%
}
\maketitle

\address{%
National Synchrotron Radiation Laboratory, University of Science and Technology of China, Hefei 230029, China
}

\begin{abstract}
This paper theoretically proves that an electron storage ring can generate coherent radiation in THz region using a quick kicker magnet and an ac sextupole magnet. When the vertical chromaticity is modulated by the ac sextupole magnet, the vertical beam collective motion excited by the kicker produces
a wavy spatial structure after a number of longitudinal oscillation periods. We calculate the radiation spectral distribution from the wavy bunch in Hefei Light Source(HLS). If we reduce electron energy to 400MeV, it can produce extremely strong coherent synchrotron radiation(CSR) at 0.115THz.
\end{abstract}




\begin{multicols}{2}

\section{Introduction}

 The synchrotron radiation will be coherent when the wavelength of the radiation and bunch length can be compared. In a storage ring, the bunch lenth is typically tens of picoseconds(ps). Therefore,the coherent radiation has a spectrum mainly in the short-wave radio and microwave regions which can be suppressed by the metallic shields \cite{PhysRev.96.180}. For ultrashort bunches, the CSR can be extended to THz range. THz wave means electromagnetic waves within band of frequencies from 0.1 to 10 terahertz. Terahertz radiation occupies a middle ground between microwaves and infrared light waves in the electromagnetic spectrum. It has many excellent features and can be widely used in research.
\par

Compared with other methods, the THz radiation from a storage ring is brilliant, broadband and stable \cite{PhysRevLett.90.094801}. In order to obtain a short-pulse electron beam bunch, the commonly used approach is quasi-isochronous operation which means the momentum compaction factor of the storage ring is reduced to 0. There are many storage ring light sources applied in this method to supply stable THz or sub-THz radiation
 \cite{PhysRevLett.90.094801,PhysRevSTAB.14.030705,PhysRevSTAB.14.040705,4441065,wep005,we5rfp010,th6pfp009} all over the world. However in these cases, the
intensity was limited by a low threshold of electron beam instability and the generation of shorter wavelength radiation required extreme stability of a ring \cite{PhysRevSTAB.11.010701}. Paper \cite{PhysRevSTAB.13.060702} proposes a more simple method for generating coherent radiation in the THz region from an electron storage ring using a quick kicker magnet and an ac sextupole magnet. When the vertical chromaticity is modulated by the ac sextupole magnet, the vertical beam collective motion excited by the kicker produces a wavy spatial structure after a number of longitudinal oscillation periods. The modulated electron bunch can produce CSR in THz region. This narrow bandwidth radiation is extremely strong, can be tuned by controlling ring parameters, and is easy to generate.
\par

This paper theoretically investigate the possibility of the above method applied in HLS, and we calculate the radiation spectral distribution from the wavy bunch in datail. Based on the results, we choose proper parameters for the kicker and the ac sextupole.

\section{Production of bunch structure}
\subsection{The bunch modulation by kicker and sextupole}
After excited by the quick kicker magnet,the vertical displacement can be expressed by:
\begin{eqnarray}
\label{yequation}
y=y_{beta}+y_{kicker}.
\end{eqnarray}
Here $y_{beta}$ is betatron motion and $y_{kicker}$ is produced by the vertical kicker magnet.
\par
The kick-angle named $\theta$, at time $t=0$ the electron bunch is excited by the kicker, then:
\begin{eqnarray}
\label{initialpos}
\begin{bmatrix}
  y \\
  y^{'} \\
\end{bmatrix}_{kicker}
=
\begin{bmatrix}
0\\
\theta\\
\end{bmatrix}
.
\end{eqnarray}
In a storage ring, the transfer matrix \cite{jinyuming} can be expressed by:
\begin{displaymath}
\begin{aligned}
\scriptscriptstyle
M(s_2/s_1)=
\begin{bmatrix}
\begin{smallmatrix}
\sqrt{\frac{\beta_2}{\beta_1}}(\cos{\Delta\psi}+\alpha_1\sin{\Delta\psi})&\sqrt{\beta_1\beta_2}\sin{\Delta\psi}\\
-\frac{(1+\alpha_1\alpha_2)\sin{\Delta\psi}+(\alpha_2-\alpha_1)\cos{\Delta\psi}}{\sqrt{\beta_1\beta_2}}&\sqrt{\frac{\beta_1}{\beta_2}}
(\cos{\Delta\psi}-\alpha_2\sin{\Delta\psi})\\
\end{smallmatrix}
\end{bmatrix}.
\end{aligned}
\end{displaymath}
Here $\alpha_1$, $\beta_1$, is Twiss parameter at position $s_1$ and $\alpha_2$, $\beta_2$ is Twiss parameter at position $s_2$. $\Delta\psi$ is the phase from $s_1$ to $s_2$.
\par
Therefore, after excited by the kicker, the displacement at any point is expressed by:
\begin{equation}
\label{kicker}
y_{kicker}=M(s_2/s_1)\begin{bmatrix}
  y \\
  y^{'} \\
\end{bmatrix}_{kicker,1}
=\theta\sqrt{\beta_1\beta_2}\sin{\Delta\psi}.
\end{equation}
Here $s_1$ is where the kicker located, namely the excited point. We only consider vertical coherent oscillation. The oscillation equation is expressed by:
\begin{equation}
\label{osciallation}
y=y_0\sin{\Delta\psi}.
\end{equation}
\par
Assuming the ac frequency is $\omega$. Then after modulated by the ac sextupole, the vertical chromaticity is expressed by:
\begin{equation}
\label{chromaticity}
\xi_y=\xi_0+\xi_1\sin{\omega t}.
\end{equation}
Here $\xi_0~and~\xi_1$ are the amplitudes of the dc (off-set) and ac (modulation) components.When $\omega=\omega_s$, $\omega_s$ is the angular frequency
of the synchrotron oscillation, the chromaticity produces a betatron phase shift \cite{PhysRevSTAB.13.060702} which is given by：
\begin{equation}
\label{omegaeqomegas}
\begin{aligned}
\delta \psi=&\frac{2\pi}{T_{rev}}\bigg[~\varepsilon(\frac{\xi_0}{\omega_s}+\frac{\xi_1}{2}t)\sin\omega_s t\\&+\tau\bigg(\xi_0\frac{\cos\omega_s t-1}{\alpha_p}+\xi_1\frac{\omega_s t \cos\omega_s t-\sin\omega_s t}{2\alpha_p}\bigg)~\bigg].
\end{aligned}
\end{equation}
Here $T_{rev}$ is electron cyclotron period, $\tau$ is longitudinal time displacement, $\varepsilon$ is energy spread and ~$\alpha_p$ is the momentum compaction factor.
\par
When $\omega_s t=n\pi,n=0,1,2\cdots$:
\begin{equation}
\label{npi}
\delta \psi=\omega_{\psi}\tau,~~~\omega_{\psi}=-\frac{2\pi}{T_{rev}}\frac{1}{\alpha_p}\bigg[(\pm 1+1)\xi_0\pm \frac{1}{2}n\pi\xi_1\bigg].
\end{equation}
It indicates that the vertical collective oscillation is nothing to do with the energy spread at these moments.In other words,the electron bunch produces a periodic structure along the longitudinal axis. The distribution of the bunch in the $\tau - y$ plane is shown in Fig. \ref{fig1}. As n is greater, the bunch repetition frequency along the longitudinal axis is higher, which produce higher CSR frequency.
\newcommand\wid{8cm}
\newcommand\hei{2cm}
\begin{center}
\includegraphics[width=\wid,height=\hei]{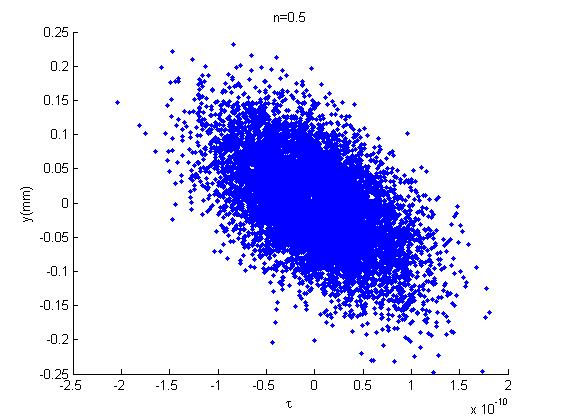}
\includegraphics[width=\wid,height=\hei]{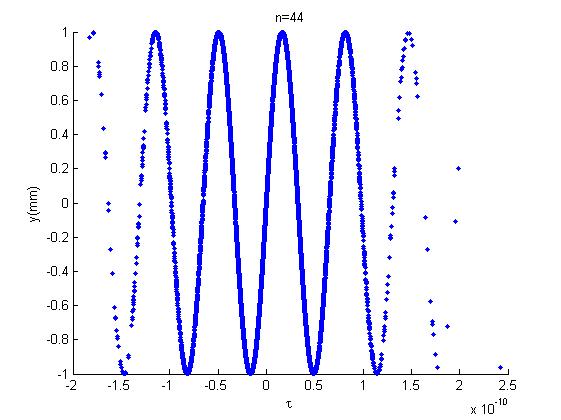}
\includegraphics[width=\wid,height=\hei]{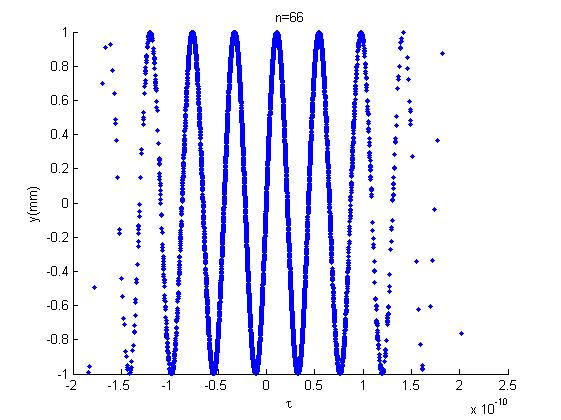}
\figcaption{\label{fig1}   At different moment, The distribution of the bunch in the $\tau - y$ plane. The phase of the reference electron is $\psi_{y0}=0$.When n is integers, the bunch produces a periodic structure along the longitudinal axis. }
\end{center}
\subsection{Radiation excitation leads to incoherent}
For convenient, we rewrite Eq.~(\ref{omegaeqomegas}) as follows:
\begin{equation}
\delta \psi=A(t)\varepsilon+B(t)\tau
\end{equation}
Assume an electron, which longitudinal parameters are $(\varepsilon,\tau)$,emits a quantum of energy $\varepsilon_N$ at time $t=t_N$. It causes the extra phase shift：
\begin{equation}
\Delta(\delta \psi)=-A(t_N)\varepsilon_N.
\end{equation}
So the vertical collective motion becomes:
\begin{equation}
\begin{aligned}
y&=y_0 \sin[\Phi_{y0}+\delta \Phi_y+\Delta(\delta \Phi_y)]\\
&=y_0 \sin(\Phi_{y0}+\delta \Phi_y)\cos[\Delta(\delta \Phi_y)]\\&+y_0 \cos (\Phi_{y0}+\delta \Phi_y)\sin[\Delta(\delta \Phi_y)].\\
\end{aligned}
\end{equation}
The expected phase shift variance \cite{PhysRevSTAB.13.060702} produced by random radiation is given by:
\begin{equation}
\label{111}
\begin{aligned}
\sigma_{\Psi N}^2
&=<\varepsilon_N^2>\int_0^tA^2(t_N)dt_N.
\end{aligned}
\end{equation}
Because $<\varepsilon_N^2>=\frac{4\sigma_\varepsilon^2}{\tau_L}$, ${\tau_L}$ is longitudinal damping time.Substituting into Eq.~(\ref{111}),we get:
\begin{equation}
\begin{aligned}
\sigma_{\Psi N}^2=&\big(\frac{2\pi}{T_{rev}}\big)^2 \frac{\sigma_\varepsilon^2}{\tau_L\omega_s^3}\Bigg\{\xi_0^2(2\omega_S t-\sin{2\omega_s t})\cos^2{\omega_s t}\\&+\xi_0\xi_1\big(\omega_s^2 t^2\cos^2{\omega_s t}-\frac{1}{4}\omega_s t\sin{4\omega_s t}\big)\\&+(\frac{\xi_1}{2})^2\big[\frac{2}{3}\omega_s^3 t^3 -\omega_s t+\sin{2\omega_s t}(\frac{1}{2}+\omega_s^2 t^2 \sin^2{\omega_s t})\big]\Bigg\}.\\
\end{aligned}
\end{equation}
Assume the extra phase shift follows the gaussian  distribution, then the bunch average position is:
\begin{equation}
\begin{aligned}
<y>&=y_0 \sin(\Phi_{y0}+\delta \Phi_y)exp(-\frac{\sigma_{\Psi N}^2}{2}).\\
\end{aligned}
\end{equation}
Where $<>$ stands for average over the extra phase shift.  As time growth, $\sigma_{\Psi N}^2$ becomes greater. The electron bunch will becomes incoherent. Meanwhile, the vertical collective oscillation amplitude becomes smaller. In principle, it requires $\omega_\psi$ as large as possible to gain higher THz CSR and $\sigma_{\Psi N}^2$~ as small as possible to maintain better coherence.

\section{Radiation from the modulation electron bunch}
\subsection{Single electron}
The far-field radiation generated by a single electron \cite{Jackson} is:
\begin{equation}
\label{sigalelectron}
\frac{d^2 I}{d\omega d\Omega}=\frac{e^2\omega^2}{4\pi^2c}\bigg|\int^{\infty}_{-\infty}\overrightarrow{n}\times(\overrightarrow{n}\times\overrightarrow{\beta})e^
{j\omega[t-\overrightarrow{n}\cdot\overrightarrow{r}(t)/c]}dt\bigg|^2.
\end{equation}
Where $\overrightarrow{n}$n is the unit vector from the electron to the observation point, $\overrightarrow{r}(t)$ stands for the position of the electron, $\overrightarrow{\beta}$ is the velocity relative to the light speed.
\par
We mark:
\begin{equation}
\label{eq:eq12}
\begin{aligned}
\overrightarrow{A}(\omega)&=\int^{\infty}_{-\infty}\overrightarrow{n}\times(\overrightarrow{n}\times\overrightarrow{\beta})e^
{j\omega[t-\overrightarrow{n}\cdot\overrightarrow{r}(t)/c]}dt\\
&\approx \frac{1}{\sqrt{3}}\bigg[-\overrightarrow{e_{\parallel}}(\frac{1}{\gamma^2}+\theta^2)K_{\frac{2}{3}}(\xi)+\overrightarrow{e_{\perp}}\theta(\frac{1}{\gamma^2}+\theta^2)
K_{\frac{1}{3}}(\xi)\bigg].\\
\end{aligned}
\end{equation}
Where
\begin{equation}
\begin{aligned}
\xi&=\frac{\omega\rho}{3c}\big(\frac{1}{\gamma^2}+\theta^2\big)^{\frac{3}{2}}.\\
\end{aligned}
\end{equation}
$\overrightarrow{e_{\parallel}}$is the unit vector in the y direction, corresponding
to horizontal polarization, and $\overrightarrow{e_{\perp}}=\overrightarrow{n}\times\overrightarrow{e_{\parallel}}.$ is the orthogonal
polarization vector corresponding approximately to vertical polarization.
Then
\begin{equation}
\frac{d^2 I}{d\omega d\Omega}=\frac{e^2\omega^2}{4\pi^2c}\bigg|\overrightarrow{A}(\omega)\bigg|^2.
\end{equation}
\par
In the electron bunch, $\overrightarrow{r}(t)$ is the reference electron position at time t, $\overrightarrow{R}$ is the vector from the reference electron to the observation point. $\theta$~ is the angle between $\overrightarrow{n}$~ and horizontal. Consider an arbitrary electron in the bunch:
\begin{equation}
\overrightarrow{r'}(t)-\overrightarrow{r}(t)=x\overrightarrow{e_x}+y\overrightarrow{e_y}+c\tau\overrightarrow{e_s}.
\end{equation}
\par
Where x, y, c$\tau$ and $\theta$ are first order small quantities. Ignoring higher-order small quantities we conclude:
\begin{equation}
\label{small}
\begin{aligned}
\overrightarrow{n}&=\cos{\theta}\overrightarrow{e_s}+\sin{\theta}\overrightarrow{e_y}.\\
\overrightarrow{n'}&=\frac{\overrightarrow{R}-\overrightarrow{r'}}{|\overrightarrow{R}-\overrightarrow{r'}|}\approx \overrightarrow{n}.\\
\overrightarrow{n^{'}}\cdot\overrightarrow{r'}(t)-\overrightarrow{n}\cdot\overrightarrow{r}(t)&=c\tau\cos{\theta}+y\sin{\theta}\approx c\tau.\\
\end{aligned}
\end{equation}
\par
So:
\begin{equation}
\label{variable}
\begin{aligned}
\overrightarrow{A'}(\omega)&=\int^{\infty}_{-\infty}\overrightarrow{n'}\times(\overrightarrow{n'}\times\overrightarrow{\beta})e^
{j\omega[t-\overrightarrow{n'}\cdot\overrightarrow{r'}(t)/c]}dt\\
&=\overrightarrow{A}(\omega)e^{-j\omega\tau}.
\end{aligned}
\end{equation}
\subsection{Coherent radiation}
The total radiation spectrum of the bunch \cite{Wiedemann} is:
\begin{equation}
\label{coherent}
\begin{aligned}
\frac{d^2I}{d\omega d\Omega}&=
\frac{e^2\omega^2}{4\pi^2c}\sum_m^{N_e}\overrightarrow{A}(\omega)e^{-j\omega\tau_m}\cdot\sum_n^{N_e}\overrightarrow{A^*}(\omega)e^{j\omega\tau_n}\\
&=\frac{e^2\omega^2}{4\pi^2c}|\overrightarrow{A}(\omega)|^2\sum_m^{N_e}\bigg[(1+\sum_{n \neq m}^{N_{e}}e^{j\omega(\tau_n-\tau_m)}\bigg]\\
&=\frac{e^2\omega^2}{4\pi^2c}|\overrightarrow{A}(\omega)|^2\bigg[N_e+\sum_m^{N_e}\sum_{n \neq m}^{N_e} e^{j\omega(\tau_n-\tau_m)}\bigg]\\
&=\frac{e^2\omega^2}{4\pi^2c}|\overrightarrow{A}(\omega)|^2\bigg[N_e+N_e(N_e-1)<e^{j\omega(\tau_n-\tau_m)}>\bigg].
\end{aligned}
\end{equation}
Where $N_e$ is the number of electrons in the bunch, $<>$ stands for average for all electrons.
\par
Assuming the electrons follow gaussian distribution in the longitudinal direction.
\begin{displaymath}
\Phi(\tau)=\frac{N_e}{\sqrt{2\pi}\sigma_\tau}exp\big(-\frac{\tau^2}{2\sigma_{\tau}^2}\big).
\end{displaymath}
Where $\sigma_\tau$ is the natural bunch length.Then
\begin{equation}
\begin{aligned}
<e^{j\omega(\tau_n-\tau_m)}>&=\int\int d\tau_n d\tau_m \frac{\Phi(\tau_n)}{N_e}\frac{\Phi(\tau_m)}{N_e} e^{j\omega(\tau_n-\tau_m)}\\
&=exp\big(-\omega^2\sigma_\tau^2\big).
\end{aligned}
\end{equation}
So the total spectrum is:
\begin{equation}
\frac{d^2I}{d\omega d\Omega}=\frac{e^2\omega^2}{4\pi^2c}|\overrightarrow{A}(\omega)|^2\big[N_e+N_e(N_e-1)exp(-\omega^2\sigma_\tau^2)\big].
\end{equation}
\subsection{Radiation from the modulated bunch}
We mark the azimuth of observation point relative to the reference electron as $\theta_0$, for an electron having vertical displacement y:
\begin{equation}
\theta=\theta_0-\frac{y}{R}
\end{equation}
Because the observation point is far enough away from the electron, $\theta$ is small quantity:
\begin{equation}
\label{A}
\overrightarrow{A}(\omega)=\overrightarrow{A_0}(\omega)+\frac{\partial\overrightarrow{A_0}(\omega)}{\partial\theta}\big(-\frac{y}{R}\big).
\end{equation}
For convenient:
\begin{equation}
\label{B}
\overrightarrow{A}(\omega)=\overrightarrow{A_0}+\overrightarrow{B_0}(-\frac{y}{R}\big).
\end{equation}
Substituting Eq.~($\ref{B}$) into Eq.~($\ref{coherent}$):
\begin{equation}
\begin{aligned}
\frac{d^2I}{d\omega d\Omega}=&
\frac{e^2\omega^2}{4\pi^2c}\sum_m^{N_e}\big(\overrightarrow{A_0}\cdot\overrightarrow{A_0^*}-\overrightarrow{A_0}\cdot\overrightarrow{B_0^*}\frac{y_m}{R}
-\overrightarrow{A_0^*}\cdot\overrightarrow{B_0}\frac{y_m}{R}\\&+\overrightarrow{B_0}\cdot\overrightarrow{B_0^*}\frac{y_m^2}{R^2}\big)
+\frac{e^2\omega^2}{4\pi^2c}\sum_m^{N_e}\sum_{n \neq m}^{N_e}\bigg[\overrightarrow{A_0}\cdot\overrightarrow{A_0^*}e^{j\omega(\tau_n-\tau_m)}
\\&-\overrightarrow{A_0}\cdot\overrightarrow{B_0^*}\frac{y_n}{R}e^{j\omega(\tau_n-\tau_m)}
-\overrightarrow{A_0^*}\cdot\overrightarrow{B_0}\frac{y_m}{R}e^{j\omega(\tau_n-\tau_m)}
\\&+\overrightarrow{B_0}\cdot\overrightarrow{B_0^*}\frac{y_m\cdot y_n}{R^2}e^{j\omega(\tau_n-\tau_m)}\bigg].
\end{aligned}
\end{equation}
Integrating from the modulated bunch:
\begin{equation}
\begin{aligned}
\label{bunch}
\frac{d^2I}{d\omega d\Omega}=&N_e\bigg\{ p_0- \frac{2y_0p_1}{R}\sin\psi_0 e^{-\frac{\omega_\psi^2\sigma_\tau^2}{2}}\\&+\frac{y_0^2p_2}{2R^2}
(1-\cos2\psi_0\ e^{-2\omega_\psi^2\sigma_\tau^2})\bigg\}\\
&+N_e(N_e-1)e^{-\omega^2\sigma_\tau^2}\bigg\{p_0\\&-p_1\frac{y_0}{R}\cdot 2\sin\psi_0\cosh\omega_\psi\omega\sigma_\tau^2 \cdot e^{-\frac{\omega_\psi^2\sigma_\tau^2}{2}}\\
&+p_2\frac{y_0^2}{R^2}\cdot\frac{\cosh2\omega_\psi\omega\sigma_\tau^2-\cos2\psi_0}{2}\cdot e^{-\omega_\psi^2\sigma_\tau^2}\bigg\}.
\end{aligned}
\end{equation}
where:
\begin{equation}
\begin{aligned}
p_0&=\frac{e^2\omega^2}{4\pi^2c}|\overrightarrow{A_0}(\omega)|^2.\\
p_1&=\frac{e^2\omega^2}{4\pi^2c}\overrightarrow{A_0}\cdot \overrightarrow{B_0^*}\\
&=\frac{e^2\omega^2}{4\pi^2c}\overrightarrow{B_0}\cdot \overrightarrow{A_0^*}.\\
p_2&=\frac{e^2\omega^2}{4\pi^2c}\overrightarrow{B_0}\cdot \overrightarrow{B_0^*}.\\
\end{aligned}
\end{equation}
\par
Where $\omega_\psi\sigma_\tau\gg1$, Eq.~($\ref{bunch}$) simplified as:
\begin{equation}
\label{simplifed}
\begin{aligned}
\frac{d^2I}{d\omega d\Omega}=&N_e p_0 \big[1+e^{-\omega^2\sigma_\tau^2}(N_e-1)\big]\\&+\frac{N_e p_2 y_0^2}{2R^2}\big[1+e^{-(\omega-\omega_\psi)^2\sigma_\tau^2}(N_e-1)/2\big].
\end{aligned}
\end{equation}
The first term of Eq.~(\ref{simplifed}) is radiation from the bunch without kicker and ac sextupole. The second term is CSR produced by the spatial structure within the bunch. The center frequency of CSR is $\omega_\psi$. We can get radiation spectrum as shown in Fig.(\ref{fig2}). At the neighbor of $\omega_\psi$, CSR is much larger than ordinary bend radiation.
\begin{center}
\includegraphics[width=8cm]{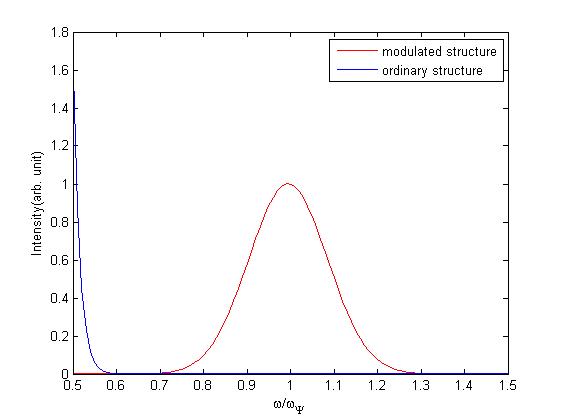}
\figcaption{\label{fig2}   The radiation spectrum produced by the modulated electron bunch.At the neighbor of $\omega_\psi$, the second term of Eq.~(\ref{simplifed}) is much larger than the first term. }
\end{center}

\section{Applied on HLS}
The required parameters of HLS \cite{HLS} are as shown in Table~\ref{tab1}.At an electron
energy of 800MeV, the nature bunch length $\sigma_\tau=50ps$, the coherent frequency(normal CSR) is 0.003THz, which lies in microwave range and can be suppressed by the metallic shields.After the deflection by the kicker magnet and modulated by ac sextupole, the center frequency of CSR is 2$\pi\times$0.026THz when $t=24\pi/\omega_s$, and $\sigma_{\Psi N}^2=1.5$. The complete results are shown in Fig. \ref{fig3}.
\begin{center}
\includegraphics[width=8cm]{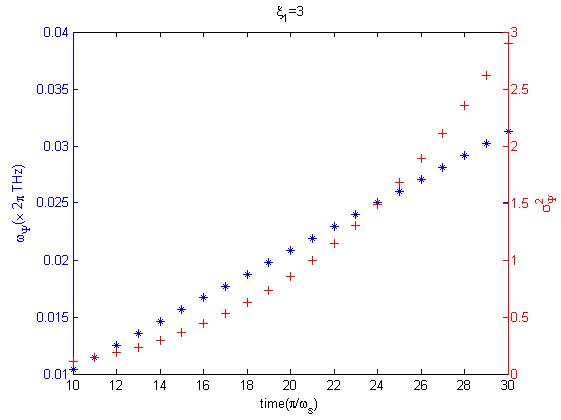}
\figcaption{\label{fig3}   When the stored electron energy is 800MeV, the center frequency and the bunch coherence varies with time. With the growth of time, the center frequency is larger while the bunch coherence is weaker. }
\end{center}
\end{multicols}

\newpage
\begin{center}
\footnotesize
\tabcaption{ \label{tab1}  Parameters of HLS.}
\begin{tabular*}{170mm}{c@{\extracolsep{\fill}}ccc}
\hline
Electron Energy&800MeV&400MeV\\
\hline
Momentum compaction factor($\alpha_P$)&0.0205&0.0205\\
\hline
Revolution frequency($f_{rev}$)&4.533MHz&4.533MHz\\
\hline
Curvature of radius of bending magnet($\rho$)&2.16451m&2.16451m\\
\hline
Natural energy spread($\sigma_\epsilon$)&0.00047&0.00024\\
\hline
Longitudinal damping time($\tau_L$)&10.8ms&86.7ms\\
\hline
Synchrotron oscillation frequency($\omega_s$)&0.193MHz&0.273MHz\\
\hline
dc chromaticity($\xi_0$)&0&0\\
\hline
ac chromaticity($\xi_1$)&3&5\\
\hline
Number of electrons in a bunch($N_e$)&$2.58\times 10^{11}$&$2.58\times 10^{11}$\\
\hline
Vertical oscillation amplitude($y_0$)&5mm&5mm\\
\hline
\bottomrule
\end{tabular*}
\end{center}

\begin{multicols}{2}
\normalsize
We can see, the CSR frequency doesn't reach THz region.If we reduce the stored electron energy to 400MeV, we can get Fig. \ref{fig4}.
\begin{center}
\includegraphics[width=8cm]{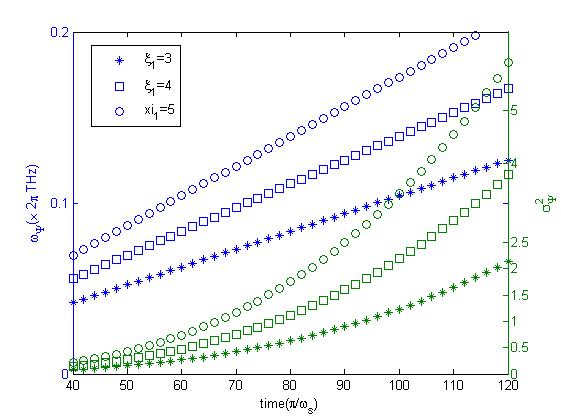}
\figcaption{\label{fig4}   When the stored electron energy is 400MeV, with different sextupole, the center frequency and the bunch coherence varies with time.  With the growth of time, the center frequency is larger while the bunch coherence is weaker.}
\end{center}
\par
From Fig \ref{fig4}, the depth of sixtupole is bigger, the center frequency of CSR can reach THz region faster,however, the bunch coherence decline faster too. Considering both respects,we choose $\xi_1=5$. $\sigma_\tau=18ps$ , $\omega_\psi=2\pi\times 0.115THz$ at $t_n=66\pi/\omega_s$~.
Further examination of Fig. \ref{fig4}, we conclude:
\begin{enumerate}
    \item At $t=0$, the bunch excited by the kicker.
    \item When $t_n=n\pi/\omega_s,n=58,59,\cdots,84$, $\omega_\psi\in [0.10,0.15]THz$, $\sigma_{\Psi N}^2\in [0.5,2]$.The bunch produces CSR in the THz region.
    \item    When $t_n \geq 120\pi/\omega_s$, $\sigma_{\Psi N}^2 \geq 5.8$, the bunch is not coherent any longer. $e^{-\sigma_{\Psi N}^2/2} \leq 5\%$, the vertical collective oscillation can be ignored.
\end{enumerate}
\par
So the frequency of the quick kicker magent satisfied:
\begin{equation}
f_{kicker}\leq\frac{\omega_s}{120\pi}=724Hz.
\end{equation}
The lower frequency can reduce repetition frequency of CSR.We hope kicker has higher frequency.
\par
Taking into account the dynamic aperture of the storage ring, at the position where the kicker located, we choose the amplitude of vertical collective oscillation $y_0=5mm$. The storage ring can tolerate this kick. $\beta_y\approx 10m$, so the kick-angle is:
\begin{equation}
\theta=\frac{y_0}{\beta_y}=0.5mrad.
\end{equation}
\par
When the stored energy is 400MeV and the ac chromaticity is $\xi_1=5$, the CSR spectrum as shown in Fig. \ref{fig5}. From the figure, FWHM=0.0138THz.
\begin{center}
\includegraphics[width=8cm]{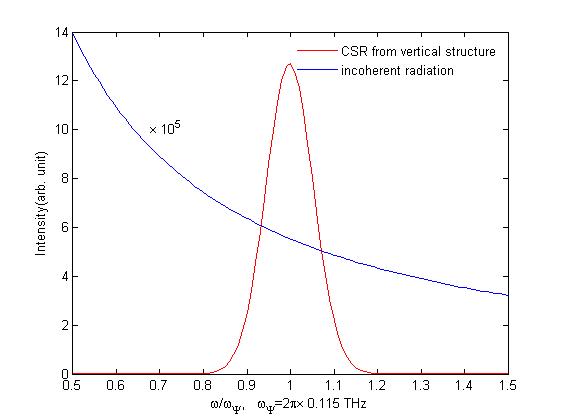}
\figcaption{\label{fig5}   When the stored energy is 400MeV, the radiation spectrum of HLS at the neighbor frequency of $\omega_\psi=2\pi\times 0.115THz$. The observation point lies in orbit plane. }
\end{center}

\section{Conclusion}
 It can produce stable, high flux CSR on HLS when the stored electron energy is 400MeV. This has important implications for expanding the scope of application of HLS. However, the CSR may cause great disturbance for storage ring. Moreover, introduction of ac sextupole can lead to nonlinear effects. It may affect the dynamic aperture of the beam. These need further research.
\end{multicols}

\vspace{-1mm}
\centerline{\rule{80mm}{0.1pt}}
\vspace{2mm}

\begin{multicols}{2}

\end{multicols}

\clearpage

\end{document}